# Hat Feed Antenna Capable of Sum and Difference Pattern Synthesis

Morteza Shahpari[(1)], Gordana K. Felic [(2)]

Centre for Wireless Monitoring and Applications, Griffith University, Australia
[(1)] morteza.shahpari@ieee.org
[(2)] National ICT Australia (NICTA), The University of Melbourne, Australia
[(2)] *Gordana.Felic@nicta.com.au*

ABSTRACT

Potentially Separate waves could propagate independently with arbitrary amplitude and phase, if one longitudinal septum is inserted in the H-Plane of the circular waveguide. The presented paper has implemented this idea in order to excite hat feed antenna simultaneously with two in-phase and out of phase signals and leads to sum and difference patterns.

## 1. INTRODUCTION

Sum and Difference patterns are essential requirements for Monopulse Tracking RADAR and Identification of Friend and Foe (IFF) Systems. Multiple antennas or Multi Mode waveguide propagation systems are main strategies to design Sum and Difference patterns. Multiple antenna designs suffer more space rather than multimode ones that leads to spillover loss in reflector systems[1-3].

In this paper, a novel structure has been proposed with simple corrections to the hat feed [4-6] antenna to achieve accomplish sum and difference patterns without extending feed shadow. Insertion of a longitudinal septum in H-Plane of hat feed and waveguide produces two split half hat antenna that are identical. As Fig.2 shows, insertion of a longitudinal septum in x-z plane (H-Plane) has leaded to transformation from a circular waveguide to two semi circle cylindrical waveguides. These identical waveguides have been used to excite semi hat antenna in front of reflector. According to the in-phase (out of phase) excitation sum (difference) pattern could be obtained. Section II deeply explains antenna geometry. Section III is dedicated to comparison of different designs. Moreover, Section IV demonstrates practical applications which could utilize proposed antennas.

## 2. ANTENNA GEOMETRY

Antenna geometry is similar to simple hat feed antenna, that is fed by circular waveguide and illuminates to paraboloid reflector. Diameter of paraboloid is $10\lambda$ and focal to diameter ratio is 0.375. Longitudinal septum saves soft surface metallic disk with radius of $1\lambda$ in front of dish and $0.66\lambda$ away waveguides.

Principal characteristics of semi circle waveguide are similar to circular waveguide and cut-off frequency, dominant and higher order modes remain unchanged. The only exception is waveguide characteristic impedance that reduces to half. Transverse electric fields in the semicircle waveguides has the same general form as in circular waveguide, hence electric field vectors for upper and lower semicircle waveguides are respectively:

$$\vec{E} = \frac{j\omega\mu}{k_c^2 r} e^{-j\beta z} J_n(k_c r)\sin\varphi \hat{u}_r + \frac{j\omega\mu}{k_c r} e^{-j\beta z} J'_n(k_c r)\cos\varphi \hat{u}_\varphi \qquad 0 < \varphi < \pi \qquad (1)$$

$$\vec{E} = \pm\frac{j\omega\mu}{k_c^2 r} e^{-j\beta z} J_n(k_c r)\sin\varphi \hat{u}_r \pm \frac{j\omega\mu}{k_c r} e^{-j\beta z} J'_n(k_c r)\cos\varphi \hat{u}_\varphi \qquad \pi < \varphi < 2\pi \qquad (2)$$

As a matter of fact if both waveguides are excited in-phase (180° out of phase), plus (minus) signs for lower waveguide will be applied and electric line forces are illustrated in Fig.2.

In the case of In-Phase excitation, field distribution on the aperture of hat antenna is in agreement with conventional hat antennas and Sum pattern will be generated. However, when waveguides are excited out of phase, reflector surface has the same amplitude distribution but experiences phase difference over $j$ variation. Difference pattern will be appeared as a result of such field distribution.

## 3. RESULTS

Antenna simulation result will be reported in this section with FEKO commercial package. Proposed antenna model has been designed to operate at 10GHz frequency. Fig.3 shows hat feed antenna radiation patterns in E plane in the sum and difference modes. Also phase of co-polarized component has been illustrated in the sum and difference modes in Fig.4. Reflector antenna radiation pattern in sum and difference modes is shown in Fig.5. The simulated gain is 26.5, 22.8dBi in sum and difference modes. It should be considered; despite the fact that simulated difference gain is much lower than maximum difference gain (2.15dB below maximum sum gain)[7], current introduced application for this antenna and some properties like; gain could be improved in the future works.

## 4. CONCLUSION

In this paper, novel configuration has been proposed to achieve sum and difference patterns. A longitudinal septum inserted in H-Plane of the hat feed antenna can create two identical independent antennas which generate sum (difference) pattern if the excitations are in-phase (out of phase).Simulation results demonstrate the feasibility of the proposed idea/method.

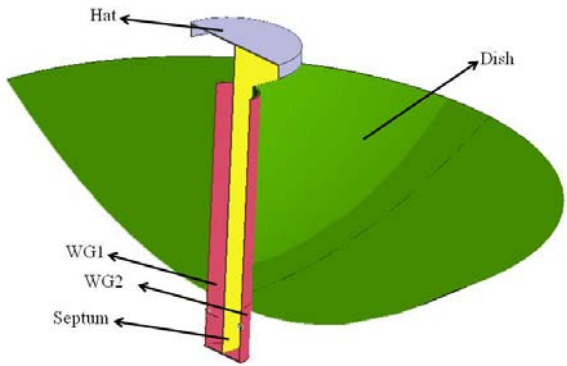

*Figure 1. Antenna Configuration*

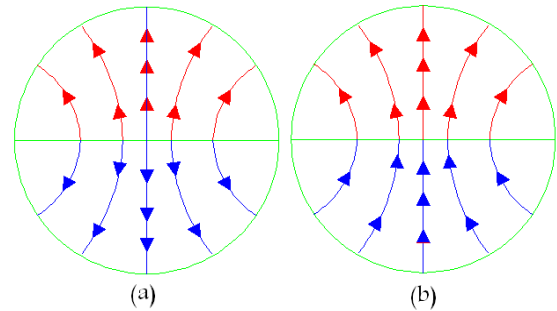

*Fig.2: Waveguide Excitation (a) Out of Phase    (b) in phase*

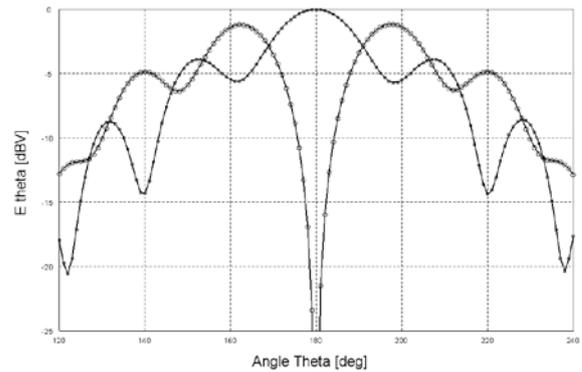

*Fig.3: Hat feed antenna radiation pattern in sum and difference modes*

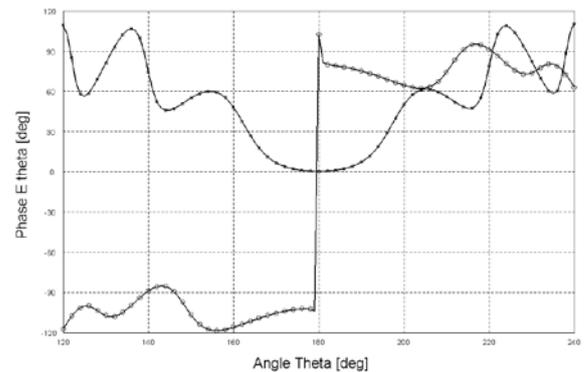

*Fig.4: phase characteristics of hat feed antenna in sum and difference modes*

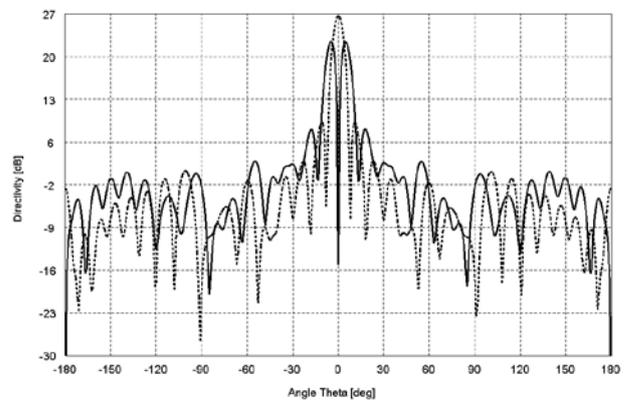

*Fig.5: secondary radiation pattern in sum and difference modes*